# Wavelength-division multiplexing communications using integrated soliton microcomb laser source

Yong Geng[1,†], Yanlan Xiao[1,†], Qingsong Bai[2†], Xinjie Han[1], Wenchan Dong[3], wenting wang[4], Jinggu Xue[5], Baicheng Yao[1], Guangwei Deng[1], Qiang Zhou[1], Kun Qiu[1], Jing Xu[3], AND Heng Zhou[1, *]

[1]Key Lab of Optical Fiber Sensing and Communication Networks, School of Information and Communication Engineering, University of Electronic Science and Technology of China, Chengdu 611731, China
[2] Chengdu Spaceon Electronics Corporation Ltd., Chengdu 610037, China
[3]Wuhan National Laboratory for Optoelectronics, Huazhong University of Science and Technology, Wuhan 430074, China
[4]Xiongan Institute of Innovation, Chinese Academy of Sciences, Hebei Province 071700, China
[5]Jiangsu Allray Inc., Zhenjiang 212000, China
*Corresponding author: zhouheng@uestc.edu.cn
†Those authors contributed equally to this Letter

**In this Letter, we investigate the feasibility and performance of wavelength division multiplexed (WDM) optical communications using an integrated dissipative Kerr soliton micro-comb as the multi-channel laser source. First, we confirm that soliton microcomb pumped directly by a DFB laser self-injection locked to the host micro-cavity has sufficiently low frequency and amplitude noises to encode advanced data formats. Second, perfect soliton crystals are exploited to boost the power level of each microcomb line, so that they can be directly used for data modulation excluding pre-amplification. Third, in a proof-of-concept experiment we demonstrate 7-channel 16-QAM data transmissions using an integrated perfect soliton microcomb as the laser carriers, excellent data receiving performances are obtained under various fiber link distances and amplifier configurations. Our study reveals that fully integrated Kerr soliton microcombs are viable and advantageous for optical data communications.**

Dissipative Kerr soliton (DKS) mode-locked frequency comb generated in optical micro-cavity has received significant research interest in recent years, thanks to its potential to transform the Nobel prize winning invention of optical frequency comb from a traditional benchtop system into an integrated or even on-chip module [1], thus to benefit wider spread application scenarios such as on-chip laser spectroscopy and metrology [2, 3], miniaturized optical clock [4] and frequency synthesizer [5], integrated microwave photonics [6], and wavelength division multiplexed (WDM) optical communications [7-10]. In particular, the impact will be huge if soliton microcombs can be practically used to replace the conventional monochromatic lasers embedded in commercial optical modules that are ubiquitous in various optical communication systems, ranging from long-haul transmission networks, data center interconnects, to 5G front haul networks. Recent studies have proven that Kerr soliton microcombs can not only provide a multitude of high-quality multi-color laser carriers [7], but also bring about profound optimizations to the basic architecture [10] and the kernel algorithms [8] of optical communication systems.

Traditionally, generating stable Kerr solitons in micro-cavities required high pump laser power and sophisticated frequency tuning schemes, so that to access the limited soliton existence range [11-13]. Therefore, Er-doped fiber amplifier (EDFA) and external cavity laser module were normally mandatory, making it infeasible to embed the whole soliton microcomb system into a practical optical module to serve as the laser carrier source. Recently, significant progress has been achieved that solitons can be reliably generated by directly pumping the micro-cavity with a distributed-feedback (DFB) laser chip, whose frequency can be well tightened to the soliton existence range within the cavity resonance via the effect of self-injection locking (SIL) [14-17]. As extra fiber amplifier and complex external cavity structure can be totally dispensed with, soliton microcombs relying on the SIL mechanism are able to achieve an unprecedented degree of simplicity and compactness, highlighted by their capability to achieve turn-key operation [14] and hybrid integration directly on silicon wafer [15]. Therefore, SIL-based fully integrated soliton microcombs constitute a much more attractive solution as a WDM laser source than their conventional counterparts.

In this letter, for the first time to our best knowledge, we conduct a specific experimental investigation on the performance matrices of using SIL-based integrated soliton microcombs for WDM optical data transmissions. Our study mainly focuses on figuring out two questions. First, whether the SIL-based soliton microcomb lines have sufficient frequency and amplitude stability to support various advanced optical data formats. Second, how to generate adequate individual comb line power for an integrated soliton microcomb to enable direct data modulation excluding extra power boost.

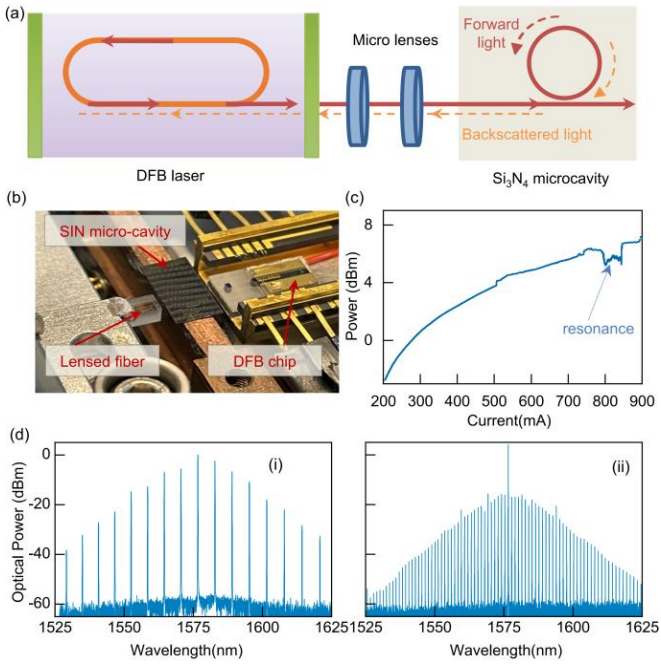

**Fig. 1.** (a) The basic principle of Kerr soliton microcomb generation pumped by a self-injection locking DFB laser. (b) The photograph of integrated Kerr soliton comb system. The high-power DFB pump laser, micro lenses, and TEC are mounted into a butterfly package. At the output side of $Si_3N_4$ chip, a lensed fiber is used to collect the soliton comb. (c) Measured power transmission of the microcavity as the DFB driving current increases from 200 mA to 900 mA. The obvious dip at about 800 mA indicates that the laser wavelength enters the micro-resonator resonance. (d-e) Measured optical spectra of 4×FSR perfect soliton crystal (i) and single soliton microcomb (ii).

The schematic of Kerr soliton microcomb generation based on direct DFB laser pumping is illustrated in Fig. 1(a), the detailed mechanism can be found in prior literature [14, 19]. The devices used in our study are shown in Fig. 1(b). A specialized high-power DFB semiconductor laser chip (Jiangsu Allray Inc) is adopted as the pump module. The laser exhibits a lasing threshold of 150 mA, the output optical power can reach 100 mW at 800 mA drive current, and the wavelength tunability is approximately 1.5 nm from 200 to 900 mA. The DFB laser is launched into a silicon nitride micro-ring resonator, which has a free spectral range (FSR) of ~150 GHz, and a loaded Q-factor of about 3 million at the pumped resonance (i.e., ~1576.6 nm). To have anomalous dispersion for dissipative Kerr soliton formation, the micro-ring waveguide is fabricated to have a cross-section of 1.65×0.8 μm$^2$ [18]. Two micro lenses are used to achieve mode profile matching and coupling loss optimization between the DFB laser beam and the inverted taper on the silicon nitride chip, and the output light from the silicon nitride chip is collected using a lensed fiber. The overall coupling loss from the DFB laser to the micro-cavity chip to the lensed fiber is about 10.0 dB. The DFB laser chip and the nitride micro-cavity chip are separately temperature controlled at the stability of ±10 mK, ensuring that the DFB laser frequency and micro-cavity resonance frequency can align with each other at the proper laser current value (i.e., with sufficient power) and not subject to severe thermal drifts.

Fig. 1(c) shows the laser power transmission recorded at the output of the micro-cavity chip, as the DFB current increases from 200 mA to 900 mA. The power drop observed at about 810 mA indicates that the DFB laser frequency encounters the micro-cavity resonance (~1576.66 nm). It is observed that the transmission lineshape exhibits a flat pit that neither resembles a Lorentz shape nor a thermal triangle, evidencing that SIL occurs between the laser and the micro-cavity [19]. Furthermore, to trigger the generation of soliton microcomb, we finely tune the current driving the DFB laser and simultaneously monitor the output optical spectrum of the micro-cavity. Importantly, under the effect of SIL, the DFB pump laser can stably access the red-detuning resonance region of the micro-cavity, as the injection locking behavior between them is much faster than the cavity thermal drift, thus we can manually tune the DFB current to access the desired dissipative Kerr soliton microcomb states. Detailed soliton formation dynamics under the effect of SIL have been elaborated in prior literature [14, 19]. Besides tuning the DFB laser current, the feedback path (i.e., SIL phase) is also critical to obtain the soliton Kerr comb state. In our experiment, the SIL phase is adjusted by precise control the distance between the $Si_3N_4$ chip and the pump laser module, using a piezo stage with a minimum step size < 30 nm.

Fig. 1(d) shows the experimentally measured optical spectra of a single soliton microcomb with 150 GHz spacing (ii) and a perfect soliton crystal with 4×FSR 600 GHz spacing (i). These two different soliton states can be reliably obtained (>50% probability for each individual manual current sweep of the DFB laser) by choosing the right temperature of the micro-cavity chip (in our experiment 34 °C for single soliton and 36.5 °C for 4×FSR perfect soliton crystal), so as to leverage the proper mode crossings and local dispersion offsets for facilitating the generation of preferred soliton microcomb states [20-21]. Importantly, it is seen that the 4×FSR perfect soliton crystal comb lines each have >10 dB higher power level than those single soliton comb lines with corresponding wavelengths, and the overall energy conversion efficiency from the pump laser to the microcomb lines is ~32% for the 4×FSR perfect soliton crystal state, in comparison, the single soliton microcomb only has ~7% energy efficiency. Superior power level and energy efficiency are critical to enable direct data modulation on the microcomb lines excluding pre-amplification, which is a prerequisite if we want to embed the microcomb sources into practical optical modules. In this sense, given the remarkable compactness of the DFB and micro-cavity chips, we believe, to serve as the WDM laser sources, it is more beneficial to interleave multiple microcombs with larger spacings than using one microcomb with smaller spacing [7]. The basic of this idea is that it is better to acquire energy directly from the pump lasers via the nonlinear parametric gain with 0 dB noise figure (NF), than referring to subsequent classic fiber amplifiers with 3 dB minimum quantum limited NF. An extra benefit of using multiple interleaved microcombs is to avoid the issue of single-point failure, which remains as a nontrivial concern for any comb-type WDM laser source. For example, if we use a single 100 GHz soliton microcomb to provide all the 40 laser carriers across the C-band, 100% channels disappear once the comb collapses; in comparison, if we interleave five 500 GHz soliton microcombs (each provides 8 channels), theoretically the overall energy efficiency is 5 times better, each individual comb line power is 25 times higher, and only 20% channels are affected when one microcomb went wrong. Of course, the interleaved comb scheme requires more DFB pump laser chips and more micro-cavities, but it is still arguably a better solution than the single microcomb scheme if we weigh up the performance, reliability, compactness, and cost.

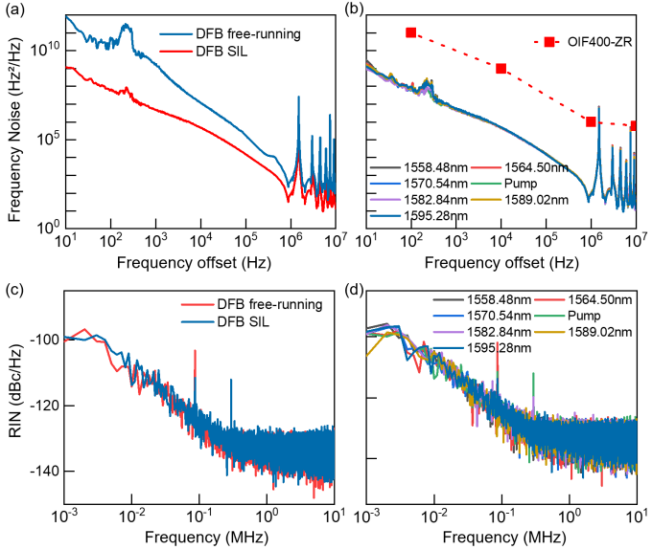

**Fig. 2.** Noise performance of integrated Kerr soliton microcomb. (a) The frequency noise spectra of free-running DFB laser (blue) and self-injection locked DFB laser (red). (b) The frequency noise spectra of 7 perfect soliton crystal comb lines ranging from 1558.5 nm to 1595.3 nm; The dashed red line is the transmitter laser FN specification defined by the OIF-400ZR-01.0 implementation agreement. (c) The RIN spectrum of free-running DFB laser (blue) and self-injection locked DFB laser (red). (d) The RIN spectrum of 7 perfect soliton crystal comb lines ranging from 1558.5 nm to 1595.3 nm.

For optical communications, laser frequency noise (FN) and relative intensity noise (RIN) are important parameters that determine the capacity and reach of the encoded optical data signals [22, 23]. Fig. 2(a-b) illustrate the FN characteristics of the DFB pump laser and the generated perfect soliton crystal microcomb lines (as shown in Fig. 1d panel i), based on the delayed self-heterodyne method [24]. It is seen from Fig. 2a that the effect of SIL indeed suppresses the free-running DFB laser frequency noise by 20~30 dB. Fig. 2b shows the FN curve for each of the soliton microcomb lines, all of which almost overlap with the pump laser FN. This phenomenon reveals an important effect for the SIL-based soliton microcomb that has not been clearly pointed out so far. In particular, it has been proved that the effect of SIL can facilitate the lock of pump-cavity detuning factor $\delta$ [19], but, the frequency drift of the DFB pump laser itself can't be fully harnessed by the micro-cavity, since under high pump power, the cavity resonances are prone to be thermally dragged by the pump frequency jitter. Mathematically, we can express the frequency jitter $\Delta\nu_m$ of the $m$-th comb line as:

$$\Delta\nu_m = \Delta\nu_p + m\Delta\nu_{rep}(\delta) \quad (1)$$

Herein $\Delta\nu_p$ denotes the frequency jitter of the pump laser, and $\Delta\nu_{rep}$ denotes the random fluctuations of the soliton repetition rate, which can be approximately considered as a function of the pump-cavity detuning factor $\delta$. For soliton microcomb directly pumped by a DFB laser, the effect of SIL can clamp the detuning factor $\delta$ and thus suppress the soliton repetition rate noise $\Delta\nu_{rep}$. However, under the condition of high pump power and thermal dynamics of the micro-cavity, the effect of SIL can't thoroughly restrain the pump laser fluctuation $\Delta\nu_p$, letting it be much larger than $\Delta\nu_{rep}$. Therefore, $\Delta\nu_m$ is predominantly set by $\Delta\nu_p$ and can become independent of the mode index $m$, as shown in Fig. 2b. Detailed investigation of this important issue will be reported elsewhere. Nevertheless, the measured FN performances of our soliton crystal microcomb lines already satisfy the criterion of typical fiber optical communication protocols. For instance, the dashed red line in Fig. 2b gives the transmitter laser FN specification defined by the OIF-400ZR-01.0 implementation agreement, and it is seen that the microcomb line FN curves are well below the required standard.

Moreover, Fig. 2(c-d) illustrate the RIN spectra of the DFB pump laser and the generated 4×FSR perfect soliton crystal comb lines. It is seen that self-injection locking hardly changes the RIN of the DFB pump laser, and the microcomb lines all inherit the RIN of the pump laser (Fig. 2d), implying that the resonance linewidth (~50 MHz) of the adopted micro-cavity does not incur too much FM-to-AM noise during SIL and soliton generation [25].

Next, we conduct optical data communication experiment by utilizing the above generated integrated soliton microcomb as the WDM laser source, the setup is shown in Fig. 3(a). At the transmitter, we utilize the 4×FSR perfect soliton crystal microcomb as the WDM source, particularly 7 comb lines ranging from 1558.5 nm to 1595.3 nm are selected as the carrier tones. The total optical power of the 7 comb lines is about +5.5 dBm and the maximum (minimum) comb line has a power of +1.5 dBm (-10 dBm), all carrier tones show high optical carrier-to-noise-ratio (OCNR) >60 dB, as shown in Fig. 3(b). The comb lines are directly (i.e., without pre-amplification) sent into a commercial optical coherent IQ modulator, which is driven by an electrical arbitrary waveform generator (eAWG) to modulate 20 Gbaud 16-QAM signal data onto all the comb lines.

After modulation, the total optical power of all data channels is about -11 dBm, which are then sent through a spool of 10 km standard single mode fiber (SSMF) to the receiver side. At the receiver, all incoming data signals are boosted together in a low-noise EDFA, and then each data channel is individually selected by a tunable optical band pass filters (OBPF) and combined with a 10 kHz linewidth local oscillator (LO) (derived from an external cavity tunable laser) for coherent data detection. The optical power of each data signal and the LO laser launched into the coherent receiver is -18 dBm and 10 dBm respectively. Subsequently, the converted baseband electrical signal is recorded by an 80 GS/s real-time digital-processing oscilloscope (DPO) and processed offline to extract the encoded data, data retrieval algorithms include chromatic dispersion equalization, carrier frequency and phase estimation, channel compensation, and timing synchronization [8].

The bit-error-ratios (BERs) of all 7-channels received data are shown in Fig. 3(c), and the constellation diagrams are also given. The BER performance for each of the data channels is well below the BER threshold of typical FEC (e.g., 7% hard FEC as shown in Fig. 3(c)). This 10 km WDM transmission result confirms that the integrated soliton crystal microcomb can be directly used as the laser source for short reach fiber links such as data-center interconnects, totally dispensing with optical amplifier at the transmitter side. Ultimately, the pre-amplifier and high-power LO laser also need to be saved by further increasing the power level of the integrated microcombs. Of, note that the total coupling loss of our microcomb apparatus is about 10 dB, mainly due to imperfect beam spot matching between the DFB, microcavity chip and lensed fiber. Such technique loss can be reduced via optimized engineering [26], and the microcomb power can thus be well increased. For long distance transmission, on the other hand, power amplifiers are unavoidable anyway to compensate for the link attenuations, and the microcomb power becomes less critical. So, in a second data transmission experiment, we add a pre-amplifier EDFA at the transmitter side (see Fig. 3(a)), and increase the fiber link to 50 km. As shown in Fig. 3c, favorable BER are again obtained for each of the 7 data channels, confirming that the intrinsic power, OCNR, FN, and RIN of the integrated soliton crystal microcomb are capable to carry high speed coherent data also for long distance transmission.

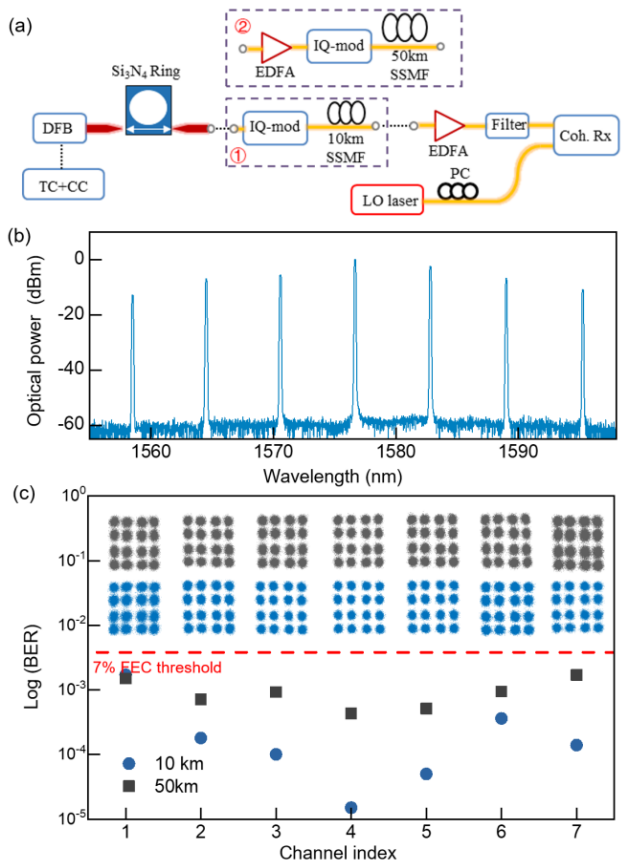

**Fig. 3.** (a) Experimental setup for WDM optical data communications using integrated Kerr soliton combs as the multi-channel laser source. TC: temperature controller. CC: current controller. PC: polarization controller. SSMF: standard single mode fiber. EDFA: Erbium Doped Optical Fiber Amplifier. (b) Measured optical spectrum of integrated perfect soliton crystal microcomb adopted as WDM laser source. (c) Measured BER for data transmissions over 10 km and 50 km SSMF links. The horizontal dashed line is the BER threshold for 7% hard FEC. Inset shows the constellation diagrams.

**Funding.** This work is supported by the National Key Research and Development Program of China (2019YFB2203103, 2021YFB-2800602), the NFSC grant (No. 62001086 and 61705033), and the Sichuan Science and Technology Program 2021YJ0095.

**References**
1. B. Stern, X. Ji, Y. Okawachi, A. L. Gaeta, and M. Lipson, Nature 562, 401 (2018).
2. N. Picqué and T. W. Hänsch, Nat. Photonics 13, 146 (2019).
3. A. Dutt, C. Joshi, X. Ji, J. Cardenas, Y. Okawachi, K. Luke, A. L. Gaeta, and M. Lipson, Sci. Adv. 4, e1701858 (2018).
4. T. E. Drake, T. C. Briles, J. R. Stone, D. T. Spencer, D. R. Carlson, D. D. Hickstein, Q. Li, D. Westly, K. Srinivasan, S. A. Diddams, and S. B. Papp, Phys. Rev. X 9, 031023 (2019).
5. W. Liang, D. Eliyahu, V. Ilchenko, A. Savchenkov, A. Matsko, D. Seidel, and L. Maleki, Nat. Commun. 6, 7957 (2015).
6. D. Marpaung, J. Yao, and J. Capmany, Nat. Photonics 13, 80 (2019).
7. P. M. Palomo, J. N. Kemal, M. Karpov, A. Kordts, J. Pfeifle, M. H. P. Pfeiffer, P. Trocha, S. Wolf, V. Brasch, M. H. Anderson, R. Rosenberger, K. Vijayan, W. Freude, T. J. Kippenberg, and C. Koos, Nature 546, 274 (2017).
8. Y. Geng, H. Zhou, X. J. Han, W. W. Cui, Q. Zhang, B. Y. Liu, G. W. Deng, Q. Zhou, and K. Qiu, Nat. Commun. 13, 1070 (2022).
9. A. Fülöp, M. Mazur, A. Lorences-Riesgo, Ó. B. Helgason, P.-H. Wang, Y. Xuan, D. E. Leaird, M. Qi, P. A. Andrekson, A. M. Weiner, and V. Torres-Company, Nat. Commun. 9, 1598 (2018).
10. M. Mazur, M. Suh, A. Fülöp, J. Schröder, V. Torres-Company, M. Karlsson, K. Vahala, and P. Andrekson, J. Lightwave Technol. 39, 4367-4373 (2021).
11. T. Herr, V. Brasch, J. D. Jost, C. Y. Wang, N. M. Kondratiev, M. L. Gorodetsky, and T. J. Kippenberg, Nat. Photonics 8, 145 (2014).
12. H. Zhou, Y. Geng, W. Cui, S. Huang, Q. Zhou, K. Qiu, and C. W. Wong, Light Sci. Appl. 8, 50 (2019).
13. X. Yi, Q. F. Yang, K. Y. Yang, and K. Vahala, Opt. Lett. 41, 2037 (2016).
14. B. Shen, L. Chang, J. Liu, H. Wang, Q.-F. Yang, C. Xiang, R. N. Wang, J. He, T. Liu, W. Xie, J. Guo, D. Kinghorn, L. Wu, Q.-X. Ji, T. J. Kippenberg, K. J. Vahala, and J. E. Bowers, Nature 582, 365 (2020).
15. C. Xiang, J. Liu, J. Guo, L. Chang, R. N. Wang, W. Weng, J. Peters, W. Xie, Z. Zhang, J. Riemensberger, J. Selvidge, T. J. Kippenberg, and J. E. Bowers, Science 373, 99 (2021).
16. N. G. Pavlov, S. Koptyaev, G. V. Lihachev, A. S. Voloshin, A. S. Gorodnitskiy, M. V. Ryabko, S. V. Polonsky, and M. L. Gorodetsky, Nat. Photonics 12, 694 (2018).
17. B. Li, W. Jin, L. Wu, L. Chang, H. Wang, B. Shen, Z. Yuan, A. Feshali, M. Paniccia, K. Vahala, and J. Bowers, Opt. Lett. 46, 5201-5204 (2021).
18. A. Kordts, M. H. P. Pfeiffer, H. Guo, V. Brasch, and T. J. Kippenberg, Opt. Lett. 41, 452 (2016).
19. A. S. Voloshin, N. M. Kondratiev, G. V. Lihachev, J. Liu, V. E. Lobanov, N. Y. Dmitriev, W. Weng, T. J. Kippenberg, and I. A. Bilenko, Nat. Commun. 12, 235 (2021).
20. D. C. Cole, E. S. Lamb, P. Del'Haye, S. A. Diddams, and S. B. Papp, Nat. Photonics 11, 671 (2017).
21. M. Karpov, M. H. Pfeiffer, H. Guo, W. Weng, J. Liu, and T. J. Kippenberg, Nat. Phys. 15, 1071 (2019).
22. E. Ip, A. P. T. Lau, D. J. F. Barros, and J. M. Kahn, Opt. Express 16, 753 (2008).
23. Y. Geng, X. Huang, W. Cui, Y. Ling, B. Xu, J. Zhang, X. Yi, B. Wu, S. Huang, K. Qiu, C. Wong, and H. Zhou, Opt. Lett. 43, 2406 (2018).
24. T. N. Huynh, L. Nguyen, and L. P. Barry, IEEE Photonics Technol. Lett. 24, 249 (2012).
25. W. Liang, V. S. Ilchenko, D. Eliyahu, E. Dale, A. A. Savchenkov, D. Seidel, A. B. Matsko, and L. Maleki, Appl. Opt. 54, 3353 (2015).
26. S. Boust, H. El Dirani, L. Youssef, Y. Robert, A. Larrue, C. Petit-Etienne, E. Vinet, S. Kerdiles, E. Pargon, M. Faugeron, M. Vallet, F. Duport, C. Sciancalepore, and F. Dijk, J. Lightwave Technol. 38, 5517 (2020).